\documentclass[conference,10pt]{IEEEtran}
\IEEEoverridecommandlockouts
\usepackage{cite}
\usepackage{amsmath,amssymb,amsfonts}
\usepackage{algorithmic}
\usepackage{graphicx}
\usepackage{textcomp}
\usepackage{multirow}
\usepackage{xcolor}
\usepackage{mathtools,amssymb}
\usepackage{comment}
\usepackage{enumitem}
\usepackage{xspace}

\usepackage{caption}
\usepackage{subcaption}

\def\BibTeX{{\rm B\kern-.05em{\sc i\kern-.025em b}\kern-.08em
    T\kern-.1667em\lower.7ex\hbox{E}\kern-.125emX}}
    
\newcommand{\bc}{\textsc{BubbleClustering}\xspace}

\newcommand{\ourmethod}{\textsc{CoRN}\xspace}

\newcommand{\random}{\textsc{RANDOM}\xspace}

\newcommand{\hideContent}[1]{}
\begin{document}

\title{Modeling and Evaluation of Clustering Patient Care into Bubbles}


\author{\IEEEauthorblockN{D.~M.~Hasibul Hasan\IEEEauthorrefmark{1},
Alex Rohwer\IEEEauthorrefmark{1}, Hankyu Jang\IEEEauthorrefmark{1},
Ted Herman\IEEEauthorrefmark{1},\\
Philip M.~Polgreen\IEEEauthorrefmark{2},
Daniel K.~Sewell\IEEEauthorrefmark{3}, Bijaya Adhikari\IEEEauthorrefmark{1},
Sriram V.~Pemmaraju\IEEEauthorrefmark{1}}
\IEEEauthorblockA{\IEEEauthorrefmark{1}\textit{Dept.~of Computer Science} 
\textit{The University of Iowa}}
\IEEEauthorblockA{\IEEEauthorrefmark{2}\textit{Dept.~of Internal Medicine} 
\textit{The University of Iowa}}
\IEEEauthorblockA{\IEEEauthorrefmark{3}\textit{Dept.~of Biostatistics} 
\textit{The University of Iowa}}
\thanks{Corresponding author: Sriram Pemmaraju (sriram-pemmaraju@uiowa.edu)}}

\IEEEaftertitletext{\centerline{(For the CDC MInD Healthcare Network)}\vspace{0.5cm}}

\maketitle

\begin{abstract}
COVID-19 has caused an enormous burden on healthcare facilities around the world. Cohorting patients and healthcare professionals (HCPs) into ``bubbles'' has been proposed as an infection-control mechanism. In this paper, 
we present a novel and flexible model for clustering patient care in healthcare facilities into bubbles in order to minimize infection spread. Our model aims to control a variety of costs to patients/residents and HCPs so as to avoid hidden, downstream adverse effects of clustering patient care. This model leads to a discrete optimization problem that we call the \bc\ problem.
This problem takes as input a temporal \textit{visit graph}, representing HCP mobility, including visits by HCPs to patient/resident rooms. The output of the problem is a \textit{rewired visit graph}, obtained by partitioning HCPs and patient rooms into bubbles and rewiring HCP visits to patient rooms so that patient-care is largely confined to the constructed bubbles. 
Even though the \bc\ problem is intractable in general, we present an integer linear programming (ILP) formulation of the problem that can be solved optimally for problem instances that arise from typical hospital units and long-term-care facilities.
We call our overall solution approach \textit{Cost-aware Rewiring of Networks} (\ourmethod).
We evaluate \ourmethod\ using fine-grained-movement data from a hospital-medical-intensive-care unit as well as two long-term-care facilities. These data were obtained using sensor systems we built and deployed.
The main takeaway from our experimental results is that it is possible to use \ourmethod\ to substantially reduce infection spread 
by cohorting patients and HCPs without sacrificing patient-care, and with minimal excess costs to HCPs in terms of time and distances traveled during a shift.
\end{abstract}


\section{Introduction}
\label{sec:intro}
There are many instances of ``bubbles'' being used to mitigate the spread of COVID-19 \cite{noauthor_support_2021}. 
These ``bubbles'' could be social, e.g., 2 or 3 families having a lot of interaction with each other, but limiting interaction with people outside the ``bubble'' \cite{nightengale_ways_2020, cnn_want_nodate}. 
``Bubbles'' are also being used on college campuses, e.g., students taking the same set of classes or students in the same dorm wing \cite{noauthor_this_nodate}. 
In this paper, we evaluate the use of ``bubbles'' in delivering patient-care in healthcare facilities.

A \textit{bubble} is simply a group of individuals who interact with each other, possibly more than usual, while limiting their interaction with individuals outside the bubble.
The goal of using bubbles is to limit the spread of an infection by making it difficult for that infection to spread beyond the bubble, \textit{while still allowing adequate interaction}. 
In a healthcare setting, there are important constraints on how bubbles can be constructed. 
If a bubble is assumed to consist of some number of patient rooms and some number of healthcare professionals (HCPs), it must be the case that the HCPs assigned to each bubble have all the skills needed to provide adequate patient-care. 
Depending on the number of bubbles being used, it may be that there are not enough physicians to assign one to each bubble and therefore we might have to leave some HCPs outside any bubble and allow interaction between these HCPs and every bubble.  
Other constraints on bubble construction could be due to available physical resources.
HCPs interact with each other outside of patient-care, e.g., in hallways and breakrooms. 
For bubbles to be successful at disease-mitigation, even these outside interactions need to be limited to HCPs within the same bubble and for this, there need to be enough separate areas to serve as breakrooms for HCPs from different bubbles.

Using bubbles may lead to downstream costs to patient-care. Using bubbles could imply less flexibility for HCPs to provide patient-care.
This could potentially lead to the patient-care load on HCPs becoming quite skewed; the volume of care that some HCPs deliver may become much larger than typical.
This could also lead to some patient needs going unmet. 

Our overall goal in this paper is to model and solve the problem of constructing bubbles in healthcare settings that mitigate infections while taking all of the above-mentioned constraints and costs into account.
Our work is motivated by the enormous burden that COVID-19 has placed on healthcare systems world wide \cite{MillerNatureMed2020}.
And within the healthcare system, long-term-care facilities have been particularly vulnerable to COVID-19 \cite{Salcher-Konrad2020.06.09.20125237}.
According to data collected by the New York Times \cite{NYT}, as of Feb 26, 2021, more than 30\% of deaths (172,000 deaths) from COVID-19 in the US are related to long-term-care facilities. 
The experimental results we present are for COVID-19 spread 
but, our techniques, results, and lessons have wider applicability.
For example, our methods also apply to healthcare-associated infections \cite{HAIsGov} caused by pathogens such as
\textit{Clostridioides difficile} (C.~diff), Gram negative multi-drug-resistant organisms, and
\textit{Methicillin-resistant Staphylococcus aureus} (MRSA), whose spread is greatly amplified in healthcare settings.

Our work contributes to a body of infection-control literature on the effectiveness of staff cohorting as a measure for reducing disease-spread. (See Section \ref{sec:related} for details.)
Our work is distinct from prior work in that we explicitly model the costs of cohorting patient-care to both patients and HCPs and aim to design cohorting strategies that are inexpensive and effective. 

\subsection{Main contributions}
We view our contributions as follows.
\begin{itemize}
	\item \textbf{Modeling:} Our first contribution is to present a flexible model for the problem of clustering patient-care in healthcare facilities so as to minimize infection spread. We do so in a manner that makes explicit a variety of costs to patients/residents (e.g., unmet demand) and to HCPs (e.g., excess load, footsteps). Our model permits a variety of modeling choices with regards to which patients/residents, HCPs, and locations participate in the clustering and it can be applied to in-patient hospital units and long-term care facilities (LTCFs) of many different types. This modeling effort leads to a discrete optimization problem that we call the \bc\ problem. 
    \item \textbf{ILP Formulation:} In general, the \bc\ problem is intractable, but we formulate it as an integer linear programming (ILP) problem that can then be solved optimally for facilities of the size we are considering (e.g., up to 60 rooms/locations, up to 40 HCPs). 
    For a given integer $K > 0$, solving the ILP yields a clustering of HCPs and locations into bubbles $B^j$, $1 \le j \le K$. Patient-care is then ``rewired'' as per these bubbles, i.e., HCPs in a bubble only deliver care at locations in that bubble.
    \item \textbf{Using fine-grained mobility data:} Our research group 
    has built and deployed sensor systems in healthcare settings for a variety of projects. 
    We use fine-grained HCP mobility data obtained from a hospital medical intensive care unit (MICU) and two LTCFs. These data have fine granularity -- of under 20 seconds -- and can accurately place HCPs in patient rooms, hallways, breakrooms, etc. Additionally, using architectural drawings of these facilities, we discretize the physical space to obtain a shortest (walking) path metric over these facilities.  
    These data serve as the basis for extensive evaluations of our methods.
    \item \textbf{Extensive evaluation using COVID-19 simulations:} 
    To evaluate our methods we perform extensive COVID-19 simulations.
    In our \textit{baseline simulations}, we overlay a COVID-19 model \cite{Jang21} on the mobility data mentioned above. We then solve the ILP formulation of the \bc\ problem to obtain $K$ bubbles and ``rewired'' interactions. We then run \textit{rewired simulations} for COVID-19 on these ``rewired'' interactions. Our results are obtained by comparing infection counts and costs of the baseline simulations versus the rewired simulations, over different numbers of bubbles and parameter settings.
\end{itemize}
We call our overall solution approach \textit{\textbf{Co}st-aware \textbf{R}ewiring of \textbf{N}etworks} (\ourmethod).

\subsection{Summary of experimental results}
In our first set of results, we show via simulations that when we use \ourmethod, COVID-19 infection counts consistently fall, both relative to the baseline and relative to a random bubble clustering.
For example, for COVID-19 simulations with $R_0 = 2.86$\footnote{$R_0$ or the \textit{basic reproduction number} is defined as the expected number of secondary cases produced by a single (typical) infection in a completely susceptible population.} and $K = 3$ bubbles in an LTCF, the mean infection count fell by 28.64\% relative to baseline and by 41.89\% relative to random bubble clustering (see Figure~\ref{fig:infection_plot_without_cost_minimization}).
Two important trends were also observed: (i) increasing the number of bubbles consistently reduced infection counts (see Figure~\ref{fig:infection_plot_without_cost_minimization}) and (ii) \ourmethod\ became relatively more effective, as infectivity of the disease increased (see Figure~\ref{fig:infection_box_plot_unbounded_cost}).

For the above-mentioned results, we use bubble clustering obtained by solving an instance of the ILP formulation of \bc\ with no bounds on costs to HCPs and patients. While this leads to relatively low unmet demand for patients, it leads to high excess load and footsteps for HCPs. We then obtain a new bubble clustering by solving an ILP instance of \bc\ in which the costs are stringently bounded.
This forces costs to substantially reduce  while increasing infections only marginally.
For example, for an LTCF for $K = 5$ bubbles, mean excess load reduced from 0.5 hrs per day per HCP to 0.204,
mean excess footsteps reduced from 518.49 m to 213.9 m (see Tables \ref{table:excess_load_and_mobility_ilp_without_cost_minimization} and \ref{table:excess_load_and_mobility_bounded_cost}),
while mean COVID-19 infection counts remained essentially unchanged (see Figure \ref{fig:comparison_bounded_ilp_and_unbounded_ilp}).
These results show that it is possible to use \ourmethod\ to substantially reduce infection spread 
by cohorting patients and HCPs without sacrificing patient-care, and with minimal excess costs to HCPs in terms of time and distances traveled.

\subsection{Related work}
\label{sec:related}
There has been work on evaluating social bubbles as a COVID-19 intervention strategy in community settings \cite{leng2020effectiveness, block2020social}.
In healthcare settings, staff cohorting is used as an infection control strategy for healthcare-associated infections such as C.~diff, MRSA, and vancomycin-resistant Enterococcus (VRE) infections \cite{austin_vancomycin-resistant_1999, price_impact_2010, jochimsen_control_1999, hussein_association_2017, abad_systematic_2020, schmidt-hieber_intensified_2007}.
In general, staff cohorting was implemented in addition to isolation of infected patients in order to strictly assign some HCPs to care for infected patients \cite{jochimsen_control_1999, price_impact_2010, schmidt-hieber_intensified_2007}.
Cohorting HCPs in this way could be costly and may not be a viable option in the shortage of HCPs during the COVID-19 pandemic \cite{needleman_nurse-staffing_2002}.
The focus of our work is mostly on bubble clustering as a preventative measure rather than as a response to a detected outbreak.

\section{Modeling the Problem}
\label{sec:problem}
\begin{figure*}[t]
    \centering
    \includegraphics[width=\textwidth]{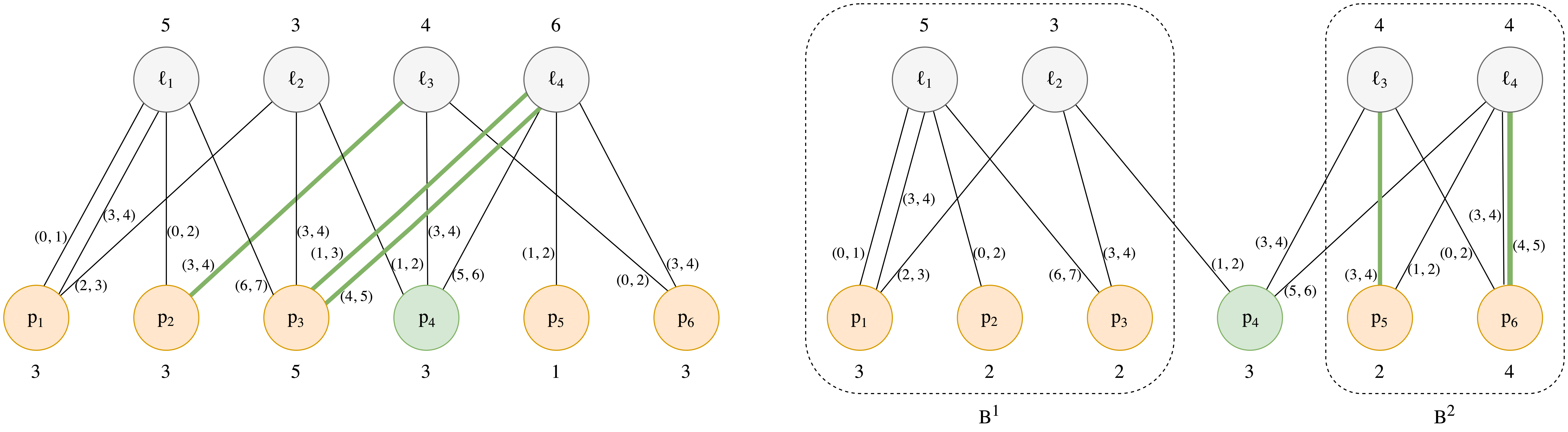}
    \caption{
    An example of a visit graph $G = (P, L, E, I)$ (left) with 
    6 HCPs. There are two types of HCPs: $P_1 = \{p_1, p_2, p_3, p_5, p_6\}$ (in orange) and $P_{ns} = \{p_4\}$ (in green) and 
    4 locations, that are all substitutable. A rewired visit graph $G_r = (P, L, E_r, I_r)$ is shown on the right.
    Numbers below the HCP nodes are their loads and numbers above the location nodes are their demands.
    Edges are labeled $(start, end)$, indicating the duration of an HCP visit to a location.
    $G_r$ is obtained by partitioning $L$ and $P_1$ into two bubbles indicated by the regions within the dashed boundaries.
    Note that $p_4$ is a non-substitutable HCP and is outside both bubbles.
    The edge $\{p_2, \ell_3\}$ and the two edges $\{p_3, \ell_4\}$ (highlighted in $G$) go between bubbles $B^1$ and $B^2$ and need to be rewired to obtain $G_r$. 
    The edge $\{p_3, \ell_4\}$ labeled $(1,3)$ in $G$ cannot be rewired, leading to unmet demand of 2 units at $\ell_4$. 
    The rewiring causes each of the two HCPs in bubble $B^2$ to have an excess load of 1 unit.}
    \label{fig:bubbleClusteringExample}
\end{figure*}

\subsection{Input}
\begin{figure*}[t]
    \centering
    \begin{subfigure}[b]{0.23\textwidth}
         \centering
         \includegraphics[width=\textwidth]{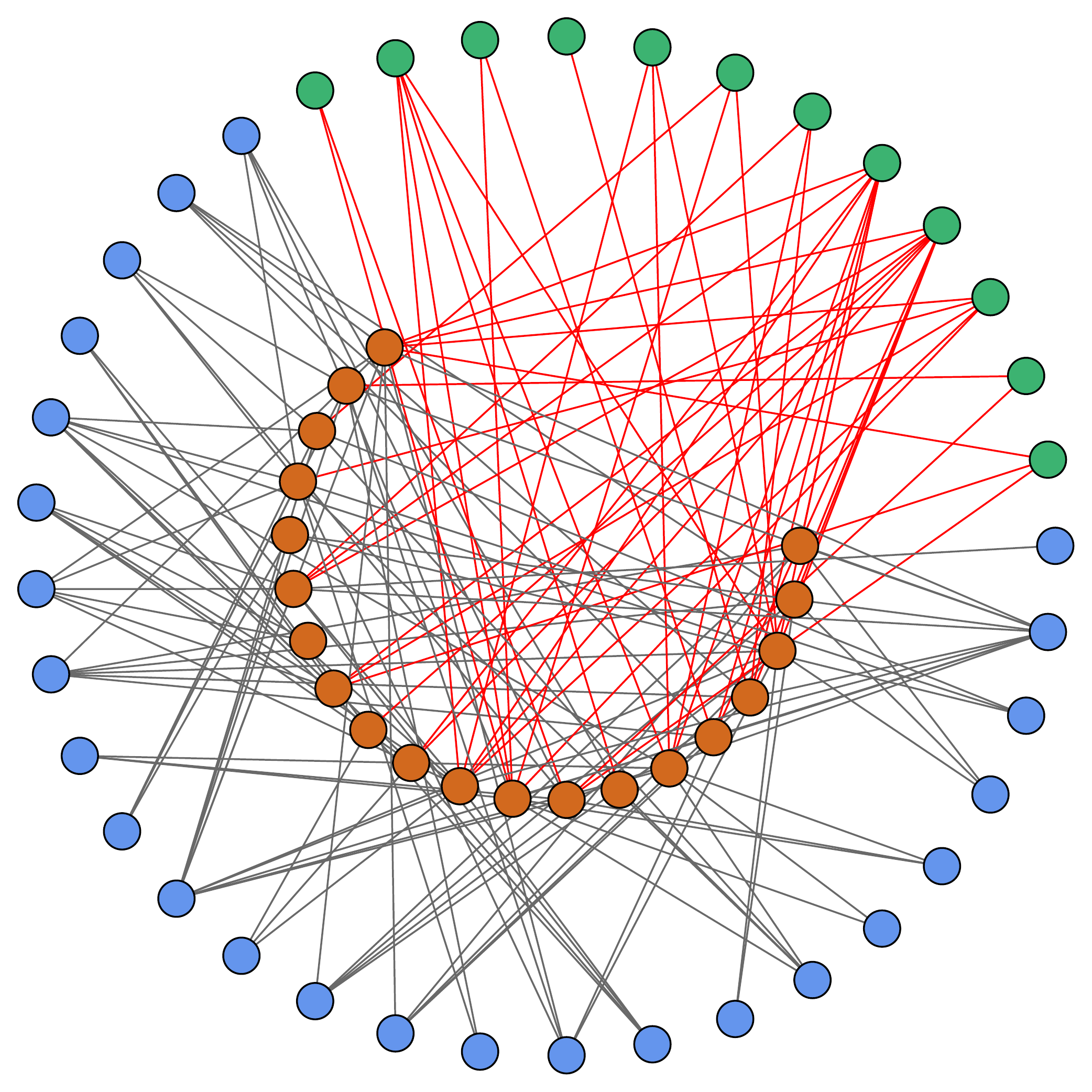}
         \caption{No bubbles}
         \label{fig:MICU_contact_network_no_bubbles}
     \end{subfigure}
     \hfill
     \begin{subfigure}[b]{0.23\textwidth}
         \centering
         \includegraphics[width=\textwidth]{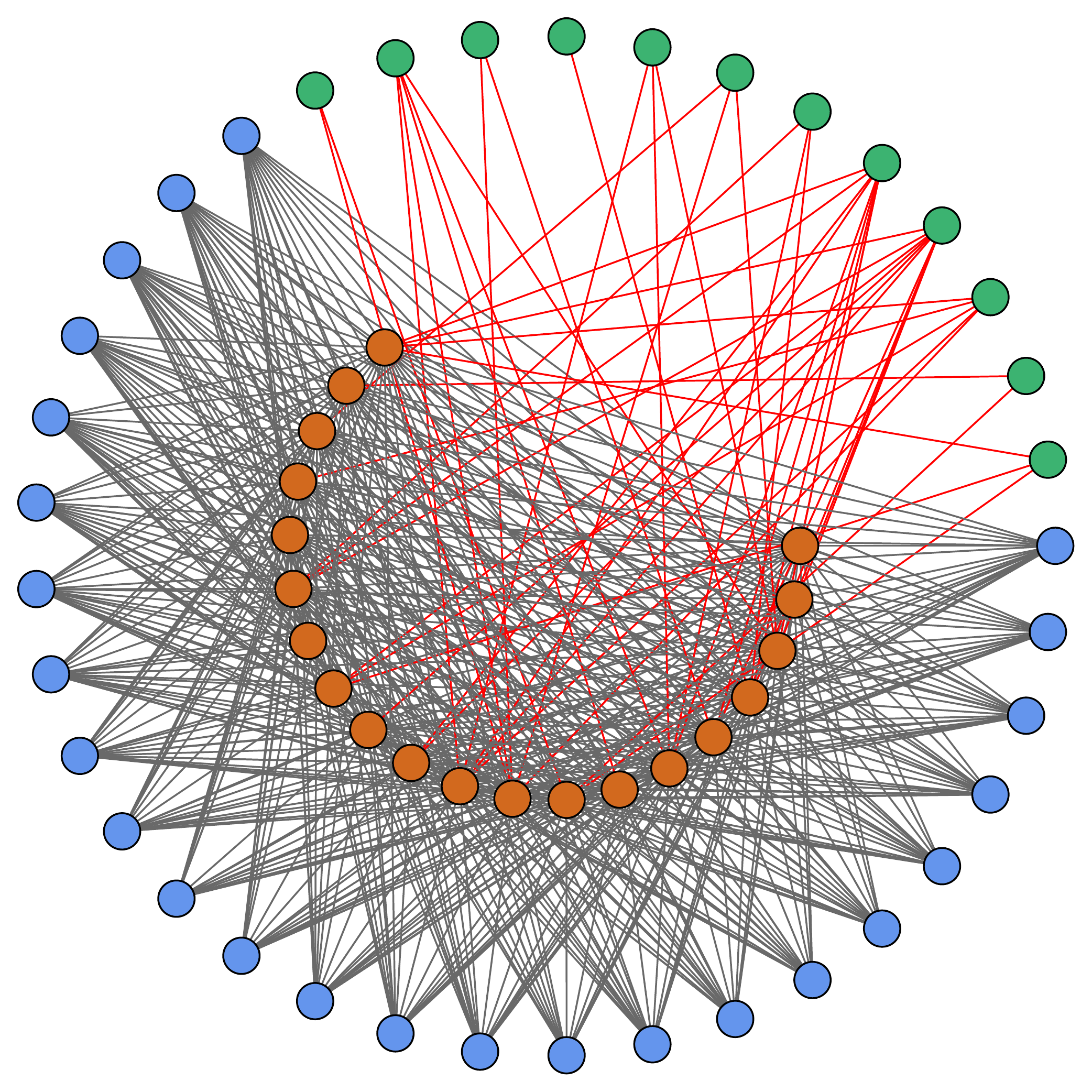}
         \caption{1 bubble}
         \label{fig:MICU_contact_network_1_bubble}
     \end{subfigure}
     \hfill
     \begin{subfigure}[b]{0.23\textwidth}
         \centering
         \includegraphics[width=\textwidth]{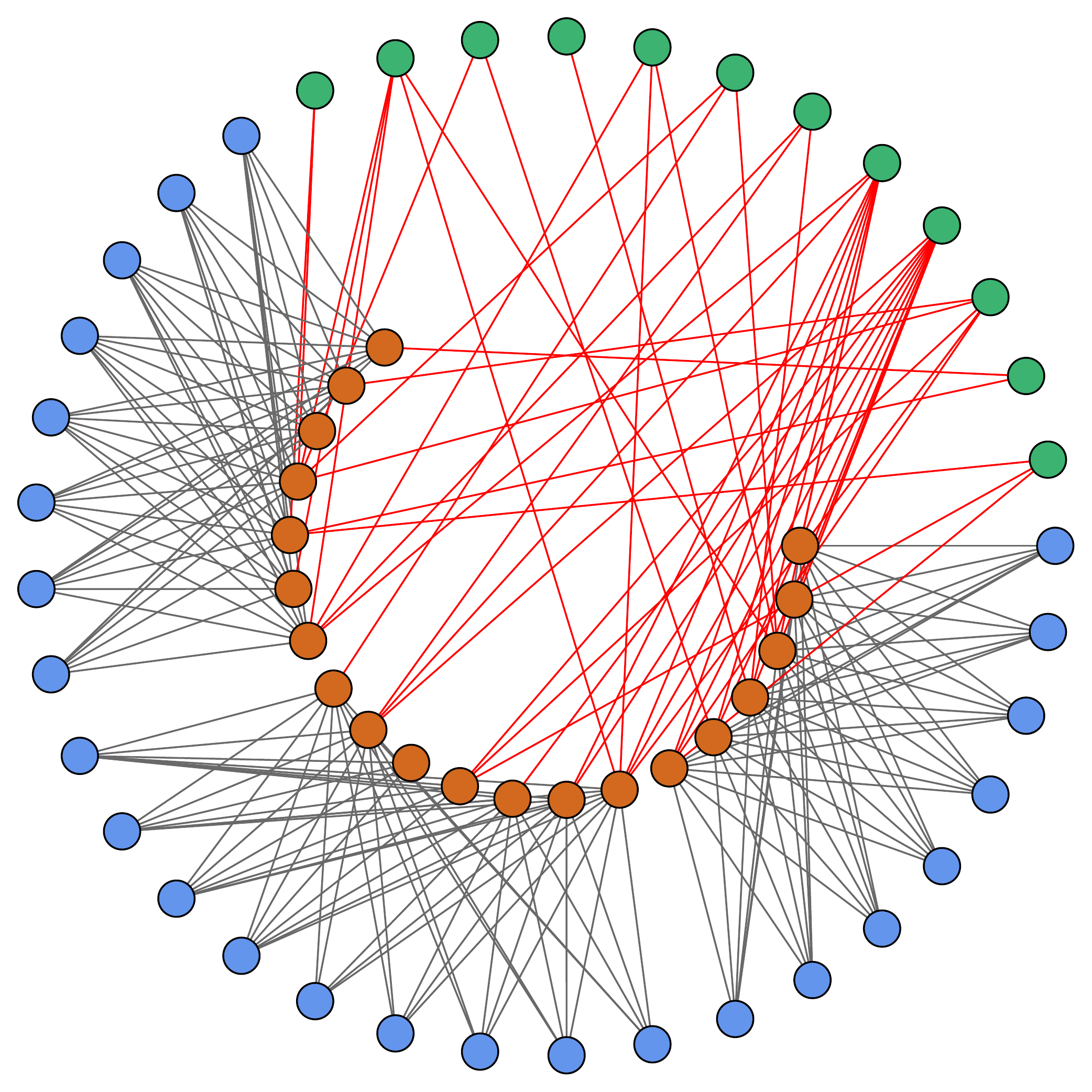}
         \caption{3 bubbles}
         \label{fig:MICU_contact_network_3_bubbles}
     \end{subfigure}
     \hfill
     \begin{subfigure}[b]{0.23\textwidth}
         \centering
         \includegraphics[width=\textwidth]{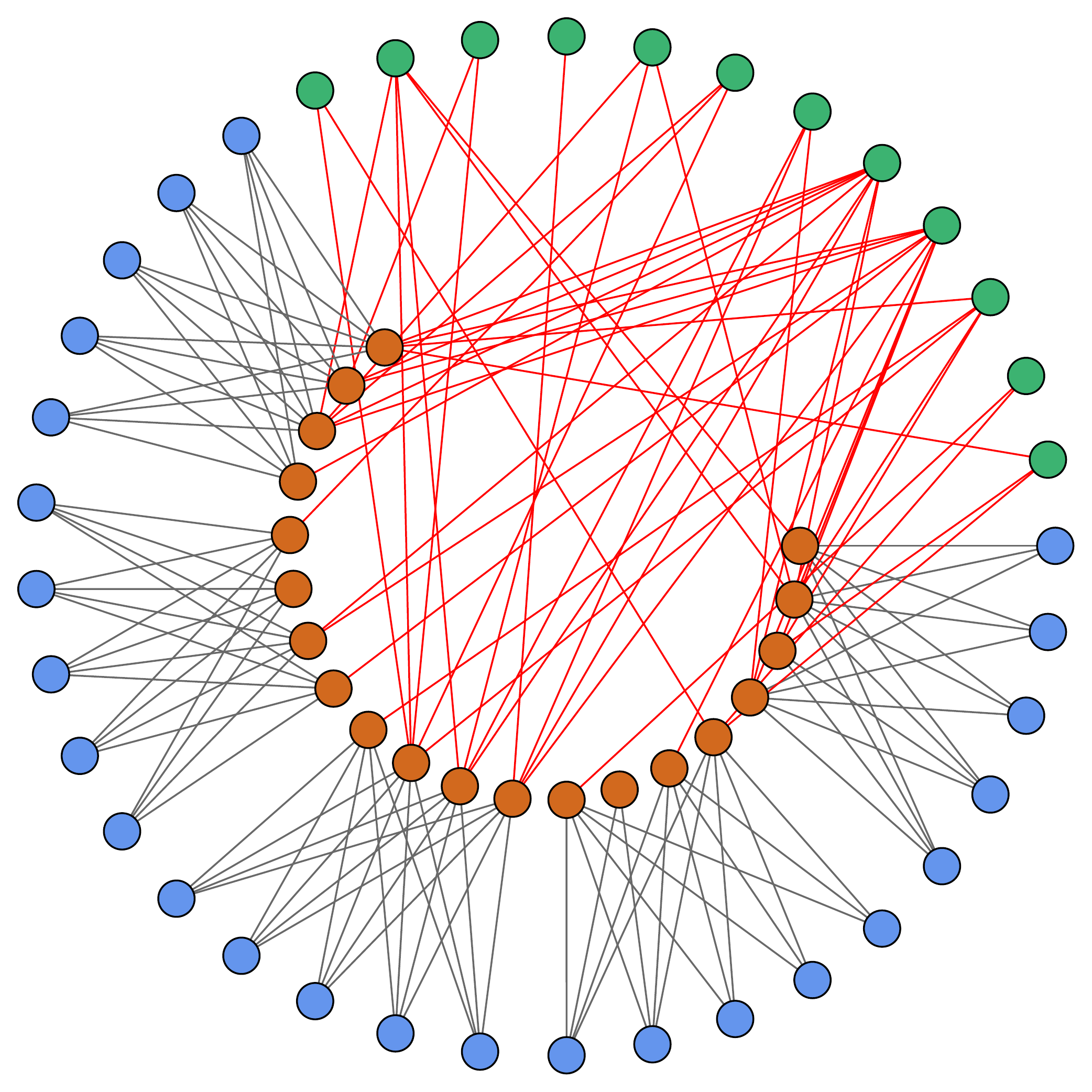}
         \caption{5 bubbles}
         \label{fig:MICU_contact_network_5_bubbles}
     \end{subfigure}
\caption{
MICU visit graph (a) and rewired visits graphs for $K \in \{1, 3, 5\}$ bubbles ((b)-(d)).
Nodes in blue, green, and brown denote nurses (25), non-nurses (15), and patient rooms (20), respectively. 
Edges in black and red denote contacts that can or cannot be rewired, respectively.
}
\label{fig:MICU_contact_network}
\end{figure*}

\begin{itemize}
    \item \textbf{Mobility log}: 
    We are given as input a \textit{mobility log} that tells us when a healthcare professional (HCP) visits a patient room and for how long. Additionally, the mobility log may also tell us about interactions between HCPs in locations that are not patient rooms, e.g., hallways, breakrooms, and nurses stations.
    We represent this mobility log as edge-labeled bipartite multigraph $G = (P, L, E, I)$ that we call a \textit{visit graph}.
    Here $P$ is the set of HCPs, $L$ is the set of locations (rooms), and $E$ is a set of edges $e = \{p, \ell\}$, denoting a visit by HCP $p \in P$ to a location $\ell \in L$. Each edge $e = \{p, \ell\}$ has a label $I(e) = (start(e), end(e))$, denoting the start and end time of the visit. We assume that for any two edges $e$ and $e'$ incident on an HCP $p$, $(start(e), end(e)) \cap (start(e'), end(e')) = \emptyset$, indicating that an HCP cannot simultaneously be in two locations. Figure \ref{fig:bubbleClusteringExample} (left) shows an example of a visit graph $G$.
    Figure \ref{fig:MICU_contact_network_no_bubbles} shows the MICU visit graph extracted from fine-grained mobility log obtained using a sensor system that we deployed.
    
    \item \textbf{HCP types}: The set $P$ of HCPs is partitioned into $H+1$ types, denoted $P_{ns}$ and $P_i$, $1 \le i \le H$. Here $P_{ns}$ denotes a set of HCPs that have no substitutes and for each $1 \le i \le H$, $P_i$ denotes a subset of HCPs that can substitute for each other. 
    Note that HCPs in $P_j$, for any $j \not= i$, cannot substitute for any of the HCPs in $P_i$. 
    For example, the HCPs at the MICU can be partitioned MICU day nurses ($P_1$), MICU night nurses ($P_2$), and a set $P_{ns}$ consisting of MICU physicians. This partitioning presupposes that the physicians visiting MICU patients have specialized skills and are therefore nonsubstitutable. MICU day nurses can all substitute for each other and similarly MICU night nurses can substitute for each, but MICU day nurses cannot substitute for MICU night nurses (and vice versa). 
    
    \item \textbf{Location types}: The set $L$ of locations is partitioned into 2 types, denoted $L_s$ and $L_{ns}$ for substitutable locations and nonsubstitutable locations, respectively. Patient (or resident) rooms would belong to $L_s$ because it is possible, by design, to substitute the visit to a patient room by an HCP $p \in P_i$ by a visit at the same time by a different HCP $p' \in P_i$. However, locations in hallways would belong to $L_{ns}$ because it is typically not possible to ``rewire'' HCP mobility to avoid hallway interactions between HCPs. Locations such as breakrooms and nurses' stations could go into either set. 
    The implication of this partitioning of HCPs and locations is that if $\{p, \ell\}$ and $\{p', \ell'\}$ are edges in the visit graph $G$ where $p, p' \in P_i$ and $\ell, \ell' \in L_s$, then we might be able to ``rewire'' these two edges, replacing them by $\{p, \ell'\}$ and $\{p', \ell\}$. 
    
    \item \textbf{Metric space on locations}: There is a metric space $D: L \times L \to \mathbb{R}^{+}$ that provides distances between pairs of locations. This metric space may be provided implicitly, as the shortest path metric of a given edge-weighted graph or explicitly as an $|L| \times |L|$ (symmetric) matrix.
    
    
    \item \textbf{Additional information on patients and residents}: We may also be given additional information on the patients or residents that the HCPs are caring for. Basic, static information consists of room occupancy (e.g., vacant, single, or double). In addition, we could be given dynamic information on when patients are discharged or transferred out, as well as when new patients arrive and into which rooms.
\end{itemize}

It is worth emphasizing the flexibility of this modeling framework. 
If the given mobility log is missing interactions in certain locations (e.g., in hallways or breakrooms) because these locations were not instrumented, it may be possible to probabilistically 
generate these interactions.
We could vary the partitioning of HCPs and locations into types depending on the healthcare facility and the policies we want to evaluate.
The metric space $D$ can be extracted from architectural drawings, as in our experiments, but it can also be quite coarse simply indicating when pairs of rooms are close and when they are far apart.
Finally, our requirements regarding additional information on patients are also quite flexible and our model can incorporate whatever information we have available.
We demonstrate the power and flexibility of our approach on two different types of facilities: a hospital MICU and stand-alone LTCFs. 

\subsection{Problem Statement}
\label{subsec:problem}

For a given positive integer $K$, our goal is to
construct $K$ bubbles, each bubble consisting of some locations and HCPs, so that care for a patient (or resident) in a certain location $\ell$ is provided by HCPs assigned to the same bubble that $\ell$ has been assigned to.
This ``rewiring'' of care so that it is clustered within bubbles aims to minimize the likelihood of infection spreading outside the bubbles.
We also want to ensure that even with this clustering of patient-care into bubbles, patients continue to receive their typical care and HCPs continue to provide their typical volume of care.

We now make this informally stated problem precise.
We start with the visit graph $G = (P, L, E, I)$ and construct a \textit{rewired visit graph} $G_r = (P, L, E_r, I_r)$ as follows.
\begin{itemize}
\item \textbf{Node clustering:} Our first step is to cluster the node set of the visit graph $G$. We partition the substitutable location set $L_s$ into $K$ non-empty, equal-sized\footnote{Since $|L_s|$ need not be a multiple of $K$, the sizes of the location bubbles may differ by at most 1.} \textit{location bubbles} $L^1, L^2, \ldots, L^K$. We partition each $P_i$, $1 \le i \le H$, into $K$ non-empty, equal-sized \textit{HCP bubbles} $P_i^1, P_i^2, \ldots, P_i^K$. This process yields $K$ \textit{bubbles} $B^j = (L^j, P_1^j, P_2^j, \ldots, P_H^j)$, $1 \le j \le K$. Note that nodes in $P_{ns}$ and $L_{ns}$ do not participate in this clustering and thus the HCPs in $P_{ns}$ and the locations in $L_{ns}$ remain outside all bubbles.

\item \textbf{Edge rewiring:} Given a set of $K$ bubbles, $B^1, B^2, \ldots, B^K$, the edge rewiring process works 
as follows.
We start by initializing the rewired edge set $E_r$ to $\emptyset$ and add rewired edges from $E$ to $E_r$ one by one. 
We consider the edges $e = \{p, \ell\} \in E$ in non-decreasing order of their start time $start(e)$.
Our goal is to add an edge incident on $\ell$ to $E_r$ with label $I_r(e) = I(e)$.
\begin{itemize}
\item[(i)] If $p \in P_{ns}$ or $\ell \in L_{ns}$, then $\{p, \ell\}$ is not rewired and is added to $E_r$ with the same label, i.e., $I_r(e) = I(e)$.
\item[(ii)] Otherwise, suppose that $p \in P_i^j$ and $\ell \in L^{j'}$.
Note that $j$ and $j'$ may be the same or distinct.
If there exists a $p' \in P_i^{j'}$ and $p'$ is \textit{available} in $G_r$ for the time period $I(\{p, \ell\})$, then add edge $\{p', \ell\}$ to $E_r$ and
set $I_r(\{p',\ell\}) = I(e)$.
If there are multiple choices for $p'$, pick one uniformly at random.
The HCP $p'$ is said to be available in $G_r$ for the  time period $I(\{p, \ell\})$ if $\cup_{\ell':\{p',\ell'\} \in E_r} I_r(\{p', \ell'\})$ does not intersect $I(\{p, \ell\})$.
Note that this definition is with respect to the set of edges currently added to $E_r$.
\end{itemize}
It is possible that some edges $\{p, \ell\} \in E$ where $p$ and $\ell$ belong to different bubbles, cannot be rewired due to lack of availability of an appropriate replacement HCP $p'$. Such edges contribute to unmet demand (defined below) at location $\ell$.
\end{itemize}
Figure \ref{fig:bubbleClusteringExample} shows an example of a visit graph $G$, a partitioning of the nodes of $G$ into two bubbles, and an edge rewiring, leading to a rewired visit graph $G_r$.
Figure \ref{fig:MICU_contact_network} shows the MICU visit graph along rewired visit graphs for $K = 1, 3, 5$.
We now make explicit the objective function we aim to minimize.
We also make explicit the costs we need to keep bounded, while minimizing this objective function.
\begin{itemize}
\item \textbf{Infection probability:} After the edge rewiring, it is guaranteed that for any edge $\{p, \ell\} \in E_r$, $p$ and $\ell$ are not in the same bubble iff $p \in P_{ns}$ or $\ell \in L_{ns}$.  
In other words, the only edges leaving a bubble are edges from a location to a non-substitutable HCP or from an HCP to a non-substitutable location. Thus any infection spread out of a bubble occurs only via non-substitutable HCPs and non-substitutable locations. Given our goal of minimizing the probability of infection leaving a bubble, we want to find a bubble clustering that yields a rewired graph in which this probability, i.e., the probability of infection traveling along edges to non-substitutable HCPs or non-substitutable locations, is minimized.
In Section \ref{subsec:transmissionWeight}, we show how to express this probability as a linear objective function. 
Our goal then is to find a bubble clustering and an associated edge rewiring that minimizes this function.

\item \textbf{Excess load:} The bubble clustering and the corresponding edge rewiring process may lead to some HCPs being assigned much more patient care than before. This excess load is an undesirable cost, not just for HCPs, but also indirectly for patients they care for.
We formalize this cost as follows.
For any HCP $p \in P$, we define the \textit{load} of HCP $p$ in graph $G$ as 
$load_G(p) = \sum_{\ell: \{p, \ell\} \in E} |I(\{p, \ell\})|$.
We define $load_{G_r}(p)$ in a similar manner, but for graph $G_r$. Then, the \textit{excess load} for an HCP $p$ is defined as 
$\max\{0, load_{G_r}(p) - load_G(p)\}$.
In the example in Figure \ref{fig:bubbleClusteringExample} the two HCPs in $B^2$ have an excess load of 1.
Our goal is to find a bubble clustering that yields a rewired graph $G_r$ that keeps the maximum excess load (over all HCPs) bounded from above.

\item \textbf{Excess footsteps:}
A cost of bubble clustering that is easy to overlook is the distance that HCPs have to travel in order to provide patient care. As shown in \cite{kharkar_hasan_polgreen_segre_sewell_pemmaraju_2020}, HCPs in a MICU (especially MICU nurses) tend to provide care to rooms that are typically close to each other.
Ideally, we want to ensure that the number of footsteps that HCPs take in order to provide care as per the rewired graph is not too much larger than the number of footsteps HCPs that take in baseline mobility data.
This is difficult to model directly and so we use the diameter of the location bubbles as a proxy for this cost. 
Define the \textit{diameter} of a bubble $B^j$ as $diam(B^j) = \max_{\ell, \ell' \in L^j}\{D(\ell, \ell')\}$.
Our goal is to find a bubble clustering that keeps the maximum bubble diameter (over all bubbles) bounded from above.

\item \textbf{Unmet demand:} After the edge rewiring, there may be edges $\{p, \ell\} \in E$ for which there is no corresponding edge in $E_r$. This is because the edge rewiring rule (iii) above may not find a replacement for edge $\{p, \ell\} \in E$ when $p$ and $\ell$ belong to different bubbles.  In this case, the care that the patient (or resident) in location $\ell$ originally received, is no longer being received. This reduction in care is a cost that we explicitly model. 
For any location $\ell \in L$, we define the \textit{demand} of location $\ell$ in graph $G$ as 
$dem_G(\ell) = \sum_{p: \{p, \ell\} \in E} |I(\{p, \ell\})|$.
We define $dem_{G_r}(p)$ in a similar manner, but for graph $G_r$. Then, the \textit{unmet demand} for a location $\ell$ is defined as 
$\max\{0, dem_{G}(p) - dem_{G_r}(p)\}$.
In the example in Figure \ref{fig:bubbleClusteringExample} location $\ell_4$ has an unmet demand of 2 units.
Our goal is to find a bubble clustering that yields a rewired graph $G_r$ that keeps the maximum unmet demand (over all locations) bounded from above.
\end{itemize}


The \bc\ problem can now be described as follows. 

\medskip

\noindent\fbox{%
    \parbox{0.97\columnwidth}{%
\textsc{BubbleClustering}\\
Given a visit graph $G = (P, L, E, I)$, a partition $(P_{ns}, P_1, P_2, \ldots, P_H)$ of the HCP set $P$, a partition $L = (L_s, L_{ns})$ of the location set $L$, and a positive integer $K$, find a bubble clustering $(B^1, B^2, \ldots, B^K)$ and the induced rewired graph $G_r = (P, L, E_r, I_r)$ that minimizes infection spread while keeping unmet demand, excess load, and excess footsteps small.
}
}

\section{Integer Linear Program Formulation}
\label{sec:ilp}
We approach the \bc\ problem by first addressing the problem of clustering locations and then the problem of clustering HCPs. We cluster locations into bounded-diameter bubbles with the aim of minimizing infection spread.
We then cluster the HCPs into bubbles with the aim of keeping excess load low.

In order to cluster the locations, we assume that there is a weight $w(\ell, \ell') \in [0, 1]$ assigned to every (unordered) pair of rooms $\ell, \ell' \in L_s$, representing the probability that infection will be transmitted from a patient in location $\ell$ to a patient in room $\ell'$ via an HCP in $P_{ns}$ or vice versa. 
Recall that $L_s$ is the set of substitutable locations, i.e., locations that participate in bubble clustering and $P_{ns}$ is the set of non-substitutable HCPs, i.e., HCPs that do not participate in bubble clustering.
Later we show how this weight assignment can be computed from the mobility log, given the transmission probability of a disease. 
With location-location weights assigned in this manner, the problem of clustering locations becomes a problem of clustering locations into $K$ equal-sized bubbles such that (i) each bubble has bounded diameter and (ii) the sum of weights $w(\ell, \ell')$ where $\ell$ and $\ell'$ belong to different bubbles, is minimized.
This problem is a generalization of the \textit{balanced minimum cut} problem \cite{AndreevRackeSPA2004,HenzingerNoeSchulzJEA2020}, in which a given edge-weighted graph is required to be vertex-partitioned into $K$ equal-sized subsets such that the edges crossing the partitions have a minimum total weight. The balanced minimum cut problem is NP-complete even in its bipartition version (i.e., when $K = 2$) \cite{GareyJohnson}. 
Furthermore, for non-constant $K$ (i.e., when $K$ is part of the input)
\cite{AndreevRackeSPA2004} show that is not possible to solve this problem in polynomial time to any finite approximation factor.
Given the intractability of this problem in general, we take advantage of the relatively small size of healthcare facilities and solve the problem optimally via an Integer Linear Program (ILP) formulation.
We then extend this ILP formulation to cluster HCPs so as to minimize excess load. 

\subsection{ILP Details}
\label{sec:ipldetails}
We start the ILP formulation by defining two 0-1 variables: (i) for $\ell, \ell' \in L_s$, variable $e_{\ell, \ell'} = 1$ iff locations $\ell$ and $\ell'$ are in different bubbles and
(ii) for $\ell \in L_s$, $1 \le k \le K$, $x_{\ell, k} = 1$ iff location $\ell$ is assigned to bubble $k$.
Given these variables, the objective function of our ILP, which we want to minimize, becomes
\begin{equation}
\label{eqn:objFunction}
\sum_{\ell, \ell' \in L_s} w(\ell, \ell') \cdot e_{\ell, \ell'}.   
\end{equation}
The constraints
\begin{align}
e_{\ell, \ell'} & \geq  x_{\ell,k} - x_{\ell',k} \qquad & \forall \ell, \ell' \in L_s, \forall k\label{eqn:connect1}\\
e_{\ell, \ell'} & \geq  x_{\ell',k} - x_{\ell,k} \qquad & \forall \ell, \ell' \in L_s, \forall k\label{eqn:connect2}\\
\sum_{k=1}^K x_{\ell,k} & =  1       \qquad & \forall \ell \in L_s\label{eqn:oneBubble}\\
\sum_{\ell \in L_s} x_{\ell,k} & \le  \left\lceil \frac{|L_s|}{K} \right\rceil \qquad & \forall k\label{eqn:equalSizes}
\end{align}
ensure that every location in $L_s$ is assigned to exactly one bubble (\ref{eqn:oneBubble}), 
every bubble is assigned the same number of locations (\ref{eqn:equalSizes}), and $e_{\ell, \ell'}$ is 1 if $\ell$ and $\ell'$ are in different bubbles (\ref{eqn:connect1},\ref{eqn:connect2}). If $\ell$ and $\ell'$ are in the same bubble, the fact that we are minimizing the objective function ensures that $e_{\ell, \ell'} = 0$.
The following constraint (\ref{eqn:diameterBound}) enforces a diameter upper bound on locations assigned to each bubble.
Since $e_{\ell, \ell'} = 0$ whenever $\ell$ and $\ell'$ are in the same bubble, this constraint ensures that every pair of locations assigned to each bubble are at most $D^*$ apart, where $D^*$ is a positive integer parameter.
\begin{equation}
\label{eqn:diameterBound}
D(\ell, \ell')\cdot (1 - e_{\ell, \ell'}) \le D^* \qquad\qquad \forall \ell, \ell' \in L_s\\
\end{equation}
This completes the portion of the ILP that clusters locations into bubbles.

In order to add constraints to this ILP that yield a clustering of the HCPs into bubbles, we introduce a 0-1 variable $z_{p,k}$, with the semantics that $z_{p,k}=1$ iff HCP $p$ is assigned to bubble $k$.
\begin{align}
\sum_{p\in P_i} z_{p,k} & \le \left\lceil\frac{|P_i|}{K}\right\rceil \qquad & \forall k\label{eqn:HCPEqual}\\
\sum_{k=1}^K z_{p,k} & = 1  \qquad & \forall p \in P \setminus P_{ns}\label{eqn:HCPExactlyOne}
\end{align}
The above two constraints ensure that HCPs $P_i$ of each type $i$ are clustered equally among the bubbles (\ref{eqn:HCPEqual}) and each substitutable HCP is assigned to exactly one bubble (\ref{eqn:HCPExactlyOne}).
Finally, we add constraints to ensure that the HCPs are clustered into bubbles in a manner that ensures that the excess load of every HCP is bounded.
To understand these constraints first note that $\sum_{p \in P_i} load_G(p) \cdot z_{p,k}$ is the total load in graph $G$ of all the HCPs of type $i$ assigned to bubble $k$.
Similarly, $\sum_{\ell \in L_s} x_{\ell,k} \cdot dem_G(\ell)$ is the total demand in graph $G$ of all the (substitutable) locations assigned to bubble $k$.
Thus the gap, $\sum_{\ell \in L_s} x_{\ell,k} \cdot dem_G(\ell) - \sum_{p \in P_i} load_G(p) \cdot z_{p,k}$ is the total excess load associated with a bubble $k$. The following constraint uses a positive integer parameter $Y^*$ to bound this gap: 
\begin{equation}
\label{eqn:boundLoad}
\sum_{\ell \in L_s} x_{\ell,k} \cdot dem_G(\ell) - \sum_{p \in P_i} load_G(p) \cdot z_{p,k} \le Y^* \qquad \forall k
\end{equation}
The objective function (\ref{eqn:objFunction}), the eight sets of constraints (\ref{eqn:connect1}-\ref{eqn:boundLoad}) described above, along with integrality constraints on the variables $e_{\ell, \ell'}$, $x_{\ell, k}$, and $z_{p, k}$ complete the ILP.  Note that this ILP is characterized by two parameters $D^*$ and $Y^*$, which respectively serve to keep excess footsteps and excess load low. For values of $D^*$ and $Y^*$ that are very small, there may be no feasible solution to the ILP.
This ILP has $O(|L|^2 + (|L|+|P|)\cdot K)$ variables and $O(|L|^2 \cdot K + |P|)$ constraints.
For our experiments over the 3 different healthcare facilities, the largest ILP instance we construct has about 850 variables and 7,200 constraints.

Note that this ILP formulation ignores unmet demand.
Unmet demand occurs when we want to rewire an edge $e = \{p, \ell\} \in E$, but there is no available HCP in $\ell$'s bubble for the time interval $I(e)$. 
Minimizing unmet demand can be formulated as a generalization of the \textit{interval partitioning} problem and it is possible to construct ILP formulations of this problem. However, the number of variables and constraints in these formulations is polynomial in the number of time intervals (visits) in our input.
The number of time intervals is quite a bit larger than the number of locations and HCPs since each HCP visits a location many times in a day.
So in the interests of keeping the ILP formulation computationally tractable, we ignore unmet demand. 
This is also partly motivated by the fact that our preliminary experimental results indicated that unmet demand in MICU was negligible. Our final results in Table \ref{table:unmet_demand_ilp_without_cost_minimization} confirm this. 

\subsection{Computing transmission weights $w(\ell, \ell')$}
\label{subsec:transmissionWeight}
For each ordered pair of locations $\ell, \ell' \in L_s$, let $P_{\ell, \ell'}$ denote the set of HCPs who have visited both locations. 
In other words, $P_{\ell, \ell'}$ is the set of all HCPs who have incident edges in $G$ to both $\ell$ and $\ell'$.
For $p \in H_{\ell, \ell'}$, let the sequence
$(e_1, e_2, e_3, \dots, e_m)$
denote the edges incident on $h$ whose other
endpoint is either $\ell$ or $\ell'$.
Without loss of generality, we assume that
the edges are ordered in increasing order of $start(e_i)$, $1 \le i \le m$.
Furthermore, without loss of generality we assume that the lengths of all visit intervals $|I(e_i)|$ are identical; if this is not the case, we can chop up the intervals to ensure this.
We now define two functions: for any $1 \le k \le m$, $pre(k)$ is the number of edges $e_i$, $i < k$, to location $\ell$ and $suf(k)$ is the number of edges $e_i$, $i > k$, to location $\ell'$. Therefore, the probability of the event $E_{\ell \ell',p}$ that infection travels from location $\ell$ via HCP $p$ to location $\ell'$ is:
\begin{equation*}
    Pr[E_{\ell \ell',p}] =\sum_{k: e_k=\{p, \ell\}} (1-z)^{pre(k)} \cdot z \cdot (1-(1-z)^{suf(k)}).
\end{equation*}
Here $z$ is the transmission probability in a time interval $I(e_i)$. Finally, we calculate the ``directed'' transmission weight $\vec{w}(\ell, \ell')$ as
\begin{equation}
\vec{w}(\ell, \ell') = 1 - \prod_{p\in H_{\ell, \ell'}} (1-Pr[E_{\ell \ell', p}]).
\end{equation}
The directed transmission weight $\vec{w}(\ell, \ell')$ is not symmetric i.e., $\vec{w}(\ell, \ell')$ need not be equal to $\vec{w}(\ell', \ell)$. This is because for infection to be transmitted from $\ell$ to $\ell'$ it requires a visit by a person to $\ell$ followed by a visit by that person to $\ell'$.
Since we want to associate a single weight in the range $[0, 1]$ to each unordered pair $\ell, \ell'$, we define 
the transmission weight $w(\ell, \ell')$ as
 $(\vec{w}(\ell, \ell') + \vec{w}(\ell', \ell))/2$.

\section{Data Description}
\label{sec:data}

\begin{figure}[t!]
    \centering
    \begin{subfigure}[b]{0.24\textwidth}
         \centering
         \includegraphics[width=\textwidth]{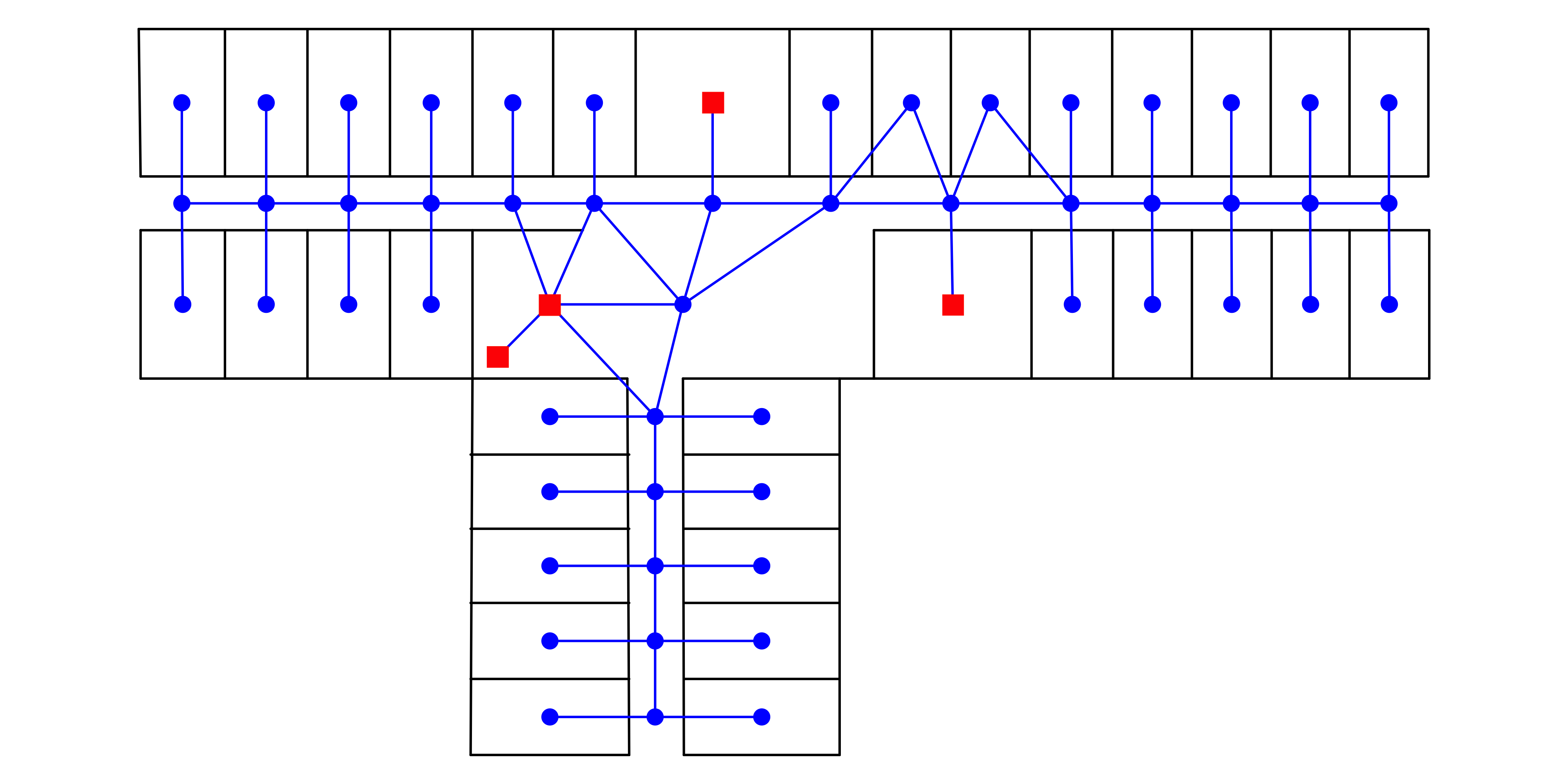}
         \caption{LTCF (small)}
         \label{fig:LTCFJuly}
     \end{subfigure}
     \hfill
     \begin{subfigure}[b]{0.24\textwidth}
         \centering
         \includegraphics[width=\textwidth]{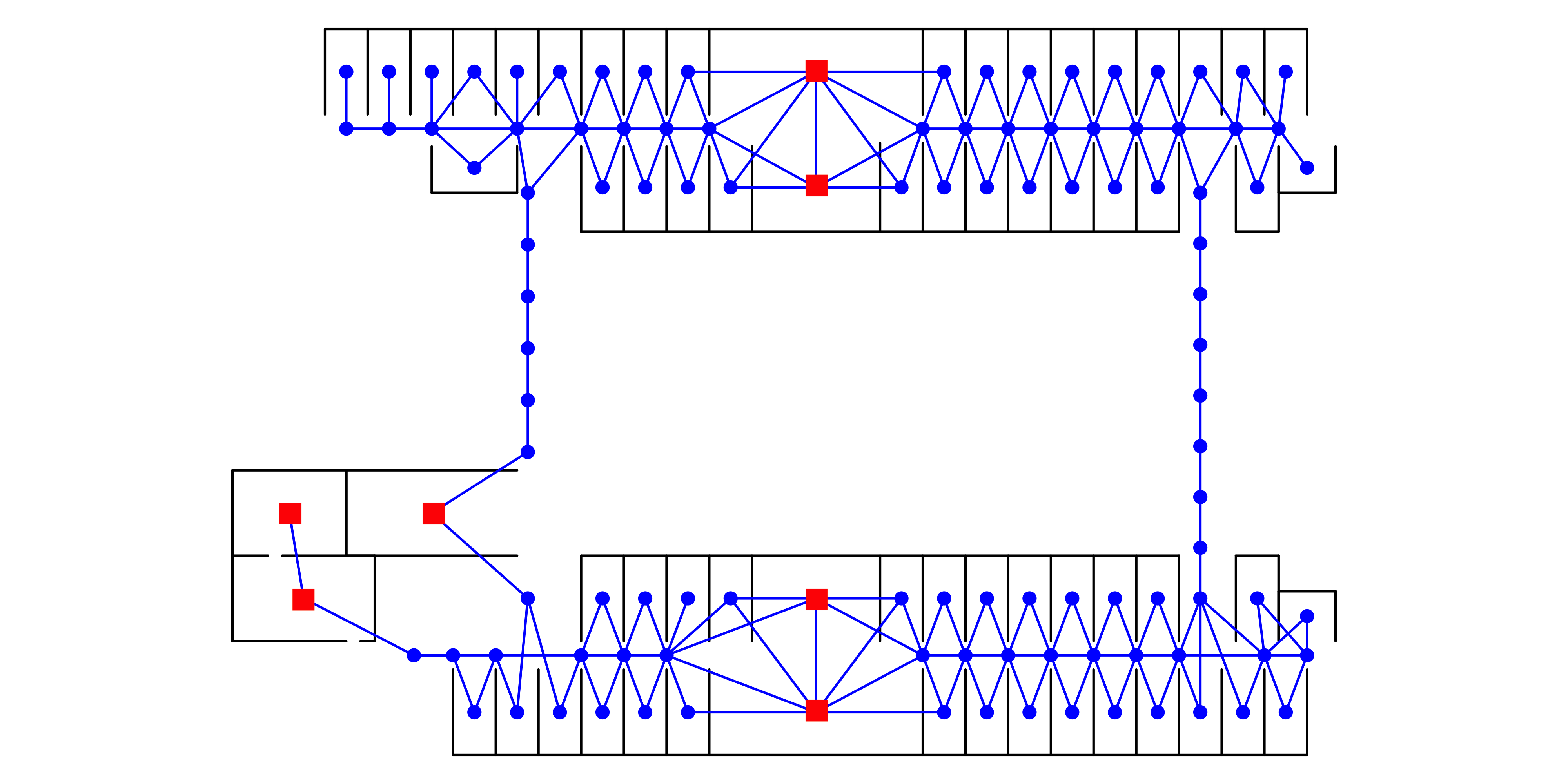}
         \caption{LTCF (large)}
         \label{fig:LTCFAugust}
     \end{subfigure}
\caption{\textbf{Spatial diagram of LTCFs}.
Spatial graphs are overlaid on LTCF architectural drawings.
Nodes are placed either in rooms (blue: resident, red: non-resident) or in hallways.
}
\label{fig:diagram_of_LTCFs}
\end{figure}


\begin{table}[b!]
\caption{Count of nurses and non-nurses (physicians, therapists, etc.) and the number of patient/resident (P/R) rooms the based on one day of observation.
}
\centering
\begin{tabular}{|c|c|c|c|}
\hline
Facility & Nurses & Non-nurses & P/R rooms \\
\hline
MICU          & 25     & 12         & 20        \\
LTCF (small)   & 8      & 6          & 33        \\
LTCF (large) & 14     & 9          & 60        \\
\hline
\end{tabular}
\label{table:hcp_and_room_counts}
\end{table}

Two types of wireless sensor devices, or \textit{motes}, are used to monitor the location of HCPs: static \textit{beacons} and mobile \textit{badges}.
Badges are distributed to HCPs at the start of their shift and collected at the end of their shift whereas beacons are placed at fixed locations.
Badges transmitted probe messages regularly -- approximately once every eight seconds in our most recent deployment. Each probe was possibly received by one or more of the beacons. The beacons recorded the time of reception along with the received signal strength indicator (RSSI) of the message, where RSSI is a number that inversely decreases as a function of distance between badge and beacon. By merging all recorded messages from all the beacons, we can estimate where each HCP was in the facility, every few seconds. 
The data used in this paper is from three deployments: a hospital MICU and two long-term-care facilities (LTCFs).
See Figure \ref{fig:diagram_of_LTCFs} for the diagram of the two LTCFs 
and Table \ref{table:hcp_and_room_counts} for the count of different types of HCPs and patient/resident rooms in each facility.

\subsection{MICU data}
We collected fine-grained movement data of HCPs in a hospital MICU over 10 consecutive days. 
Beacons were placed near patient beds, hallways, and walls while badges were distributed to HCPs.
On a single day, there were 37 badged HCPs in the facility:
25 nurses (15 day nurses and 10 night nurses) and 
12 non-nurses (doctors).
From the mote data, we extracted data on HCP visits (start and end times) to patient rooms in MICU.
Note that because of specifics of the mote deployment, these data do not contain HCP mobility when HCPs are outside patient rooms. 
We have also constructed a ``hospital graph'' that is a discretization of the physical space of the MICU, 
allowing us to impose a metric space $D$ over the set of MICU patient rooms.

We use one day of HCP mobility data and repeat it
for 30 consecutive days.
The reason for the repetitive usage of one-day of HCP mobility is that due to privacy reasons we were not allowed to track HCPs over time (badge distribution was at random each day).
Separately, we retrieved MICU patient room occupancy from anonymized admission-discharge-transfer records of patients in and constructed synthetic patient admission and discharge for a 30-day period by sampling from this data.
As we do not have HCP mobility data outside patient rooms in the MICU, we use random mixing to place HCPs outside patient rooms.
In Figure \ref{fig:MICU_contact_network} (a), we visualize the
MICU visit graph obtained from these data. 

\subsection{Small LTCF} 
In 2019, we collected HCP mobility data for the three consecutive days at an LTCF (with 33 resident rooms) in Georgia.
Beacons were placed in resident rooms while badges were distributed to HCPs.
On a single day, there were 14 HCPs in the facility:
8 nurses (5 Certified Nursing Assistants (CNAs) and 3 Licensed Practical Nurses (LPNs)) and 6 non-nurses (e.g., therapists and nutritional specialists).
We discretized the architectural map of the LTCF into a spatial graph with 57 nodes and 63 edges:
each node in the spatial graph is either a room or a location in the hallway.
We did not have data on the times when resident rooms were occupied, so assumed that each resident room was occupied by a resident all the time.
As for the MICU, we selected one day of HCP mobility data and repeated this over 30 consecutive days to generate a visit graph.

\subsection{Large LTCF}
In 2019, we also collected HCP mobility for three consecutive days at another LTCF with 60 resident rooms.
Beacons were placed in resident rooms, nursing areas, common areas, dining rooms, and therapy rooms and badges were distributed to HCPs.
On a single day, there were 23 HCPs in the facility:
14 nurses (9 CNAs and 5 LPNs) and 
9 non-nurses (e.g., therapists).
We discretized the map of this LTCF into a spatial graph with 115 nodes and 237 edges.
The rest of the procedure is similar to how we processed data from the small LTCF facility.

\subsection{Statement on data collection}

We did not gather any patient data from our instrumentation of the MICU or the LTCFs.  All HCP mobility data that we gathered was completely anonymous. Badges were arbitrarily distributed to HCPs at the start of each day, ensuring that there was no necessary connection between HCPs with the same badge ID across different days.  
The MICU data collection was determined to not be human subjects research under IRB 201308734.
The LTCF data collection was covered under IRB 201904806.

\section{Simulation Setup}
\label{sec:experiment}

    \begin{figure*}[t!]
        \begin{tabular}{c}
        \centerline{\includegraphics[width=1\textwidth]{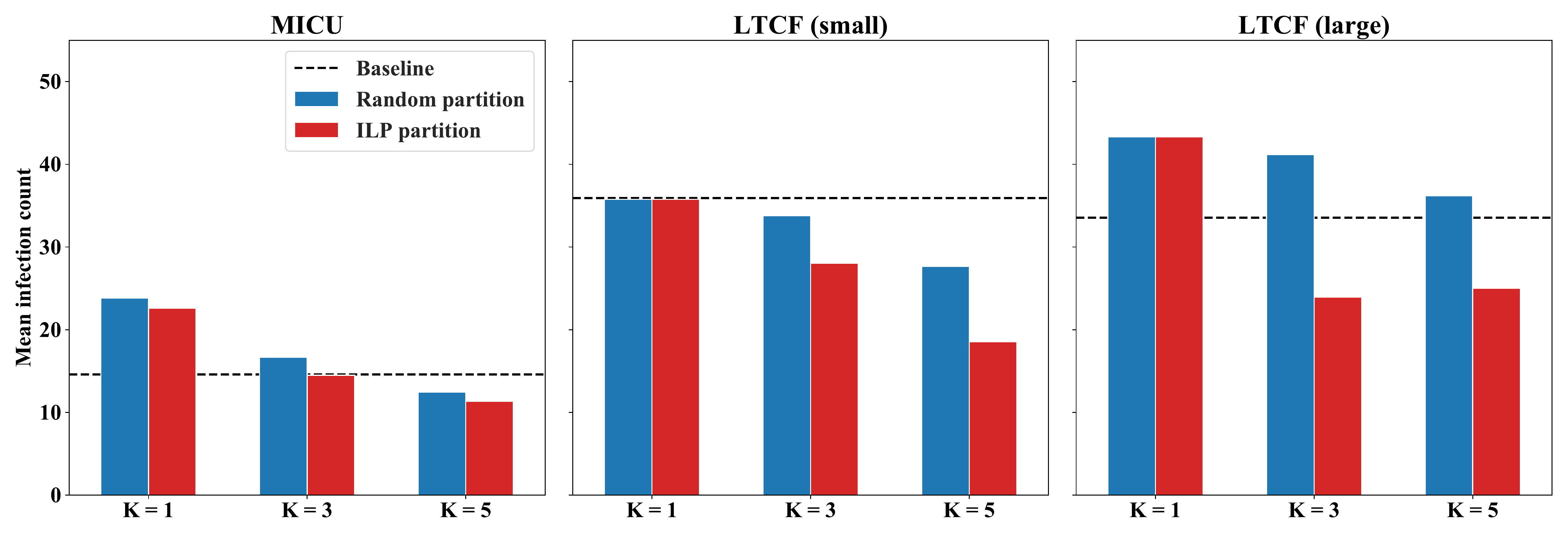}} \\
        \end{tabular}
        \caption{
        COVID-19 mean infection counts across different bubbles for different facilities for infectivity $\rho=10 \times 10^{-4}$ ($R_0 = 2.86$) over 500 replicates. \ourmethod comprehensively outperforms \random.
        Infection counts for \ourmethod\ are also below the baseline for $K > 1$, substantially so for the LTCFs.
        }
        \label{fig:infection_plot_without_cost_minimization}
    \end{figure*}


We evaluate \ourmethod\ by simulating COVID-19 spread in the MICU and in the two LTCFs that our data comes from.
In our COVID-19 model, the level of infectiousness of an infected agent ramps up exponentially, reaches maximum level on the day when symptoms start, and then ramps down exponentially, as described in detail in~\cite{Jang21}.
This aspect of the model is motivated by the fact that such temporal dynamics of viral infectivity in SARS-CoV-2 have been observed in COVID-19 patients \cite{he2020temporal, wolfel2020virological, zhou2020viral, to2020temporal}.
The level of infectiousness of an infected agent is denoted by a parameter $\beta$ whose value depends on the number of days since the patient was infected. 
For an interaction of duration $d$ between an infected agent with level of infectiousness $\beta$ and a susceptible agent, there is a probability $\rho \cdot d \cdot \beta$ of the susceptible agent getting infected via this particular interaction.
Here $\rho$ is a base infectivity that is chosen in our experiments to ensure that the $R_0$ value is in the accepted range of 2.85 and 3.75~\cite{chinazzi_effect_2020, li_early_2020, munayco_early_2020, salje_estimating_2020, SancheEID2020, ke_fast_2020}. 

Unless otherwise mentioned, we repeat the simulation 500 times (each repetition is termed a \textit{replicate} in what follows) for each setup, starting with an infected nurse chosen uniformly at random. For each replicate we compute the total number of infections observed and after 500 replicates, which we run in parallel, we compute various statistics of the total number of observed infections.

\par\noindent \textbf{\bf Baseline simulations:} 
In our \textit{baseline simulations} we run the above-described COVID-19 simulations on the visit graph $G$ obtained from the original input mobility log. 
Note that this means that we simply replay the interactions (visits) encoded in the edges of $G$ and let infections spread stochastically among patients/residents and HCPs.

\par \noindent \textbf{\bf Rewired simulations:} Once we cluster the visit graph $G$, into $K$ bubbles $\{B^j| 1\leq j \leq K \}$ and obtain the rewired graph $G_r$, using our method \ourmethod, we re-run the simulations on $G_r$.
As for the baseline simulations, we compute statistics on the total observed infection counts in the \textit{rewired simulations}.
As in the baseline simulations, the rewired simulations also start with a seed infected nurse, chosen uniformly at random. 
We make two other noteworthy changes to the rewired simulations. First, giving that we are rewiring visits by HCPs to patient/resident rooms, HCP visits to other locations (e.g., hallways) may no longer be feasible. In other words, an HCP $p$ may have spent time interval $[t, t']$, $t' > t$, at a hallway location in $G$, but in $G_r$ HCP $p$ may be spending a portion of $[t, t']$ in a patient/resident room. Hence, in $G_r$, HCP $p$ cannot continue to spend time interval $[t, t']$ in a hallway location.
So we randomly generate HCP visits to locations that are not patient/resident rooms by taking into account their availability. Second, 
HCPs who are in different bubbles are less likely to interact in casual settings (e.g., hallways). So we introduce a scaling factor of 0.75, and retain each interaction in a location that is not a patient/resident room between HCPs who are in different bubbles with probability $0.75$.
   
\par \noindent {\bf Random bubble clustering:} 
In addition to comparing rewired simulations against baseline simulations, we also compare them against simulations performed on a randomly rewired visit graph.
More precisely, given the visit graph $G = (P, L, E, I)$ and the parameter $K > 0$ representing the number of bubbles we want, we first create $K$ equal-sized \textit{random} clusters of substitutable HCPs $P_i$'s and substitutable locations $L_s$. Then we rewire the edges of $G$, as described in Section \ref{sec:problem}. 
We run the above-described COVID-19 simulations on this randomly rewired visit graph.
In what follows, we use \random refer to this approach.

\section{Results and Discussion}
\label{sec:results}

\begin{figure}[]
    \centering
    \includegraphics[width=0.5\textwidth]{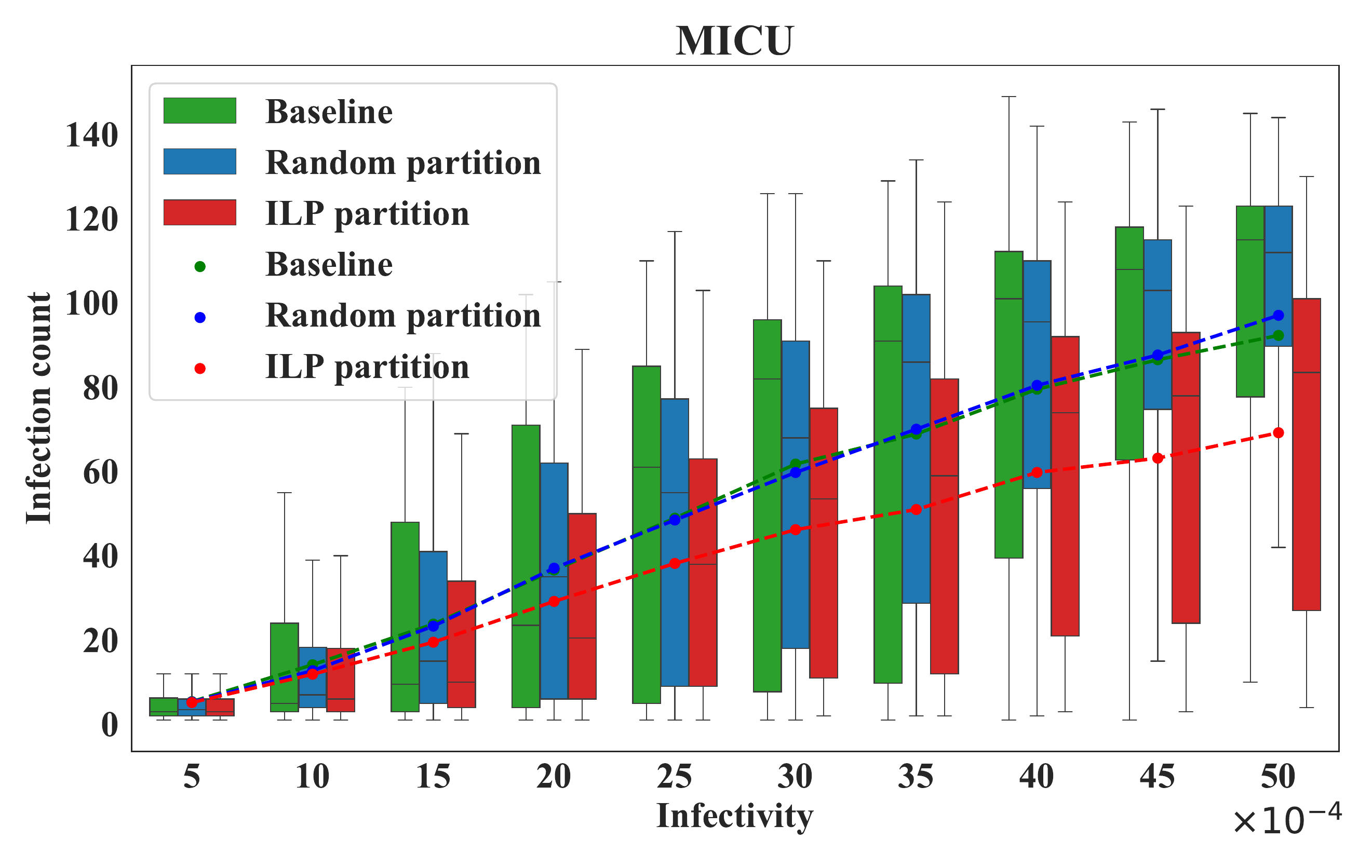}
    \caption{
    Infection counts (dashed line: mean) for \ourmethod with $K=5$ for different infectivities $\rho$, yielding $R_0 \in [1.67-7.9]$.
    }
    \label{fig:infection_box_plot_unbounded_cost}
\end{figure}

The goal of our experiment is to demonstrate that our \bc problem formulation is sound and our approach \ourmethod is successful in mitigating the infection flow. Specifically, our experiments were designed to answer the following questions.
\begin{itemize}
    \item[(i)] Does the bubble clustering generated by \ourmethod\ reduce infection?
    \item[(ii)] How does infection-spread change as the number of bubbles $K$ used by \ourmethod increases and how does the effectiveness of bubble clustering change as the infectivity $\rho$ increases?
    \item[(iii)] Are the parameters $D^*$ and $Y^*$ which upper bound excess footsteps and excess load really necessary? In other words, could we obtain low-cost bubble clustering even with $D^* = Y^* = \infty$? 
    \item[(iv)] On the other hand, if we do need to use small values of 
    $D^*$ and $Y^*$ to force costs to be low, does this cause infections to increase unacceptably?
\end{itemize}

We describe our experiments and results for each of these questions next.

\subsection{Infection minimization}
\label{subsec:infmin}

In our first set of experiments, we measure the effectiveness of the bubbles generated by \ourmethod 
relative to the baseline and \random in mitigating infection flow. 
Our results is presented in Figure \ref{fig:infection_plot_without_cost_minimization}. 
The horizontal line represents the mean COVID-19 infection count in the baseline simulation and the bars represent the same for 
\ourmethod\ and \random.
As shown in the figure, as the value of $K$ increases the mean infection count decreases for bubble clustering methods as expected. The next observation we make is that our approach outperforms \random consistently across datasets and for all values of $K$ by up to 41.89\%. This is because \random does not consider the infection spreading via nodes in $P_{ns}$ while determining bubbles. Our results imply that not all bubbles of the same size are equal when it comes to infection control and explicitly minimizing the probability of infection flow between locations leads to effective bubbles. 

We see that for $K = 1$, the baseline simulation has a lower mean infection count than \ourmethod. The reason behind this is that  nurses in the baseline visit graph $G$ naturally cohort with each other in their mobility patterns, limiting their susceptibility to infection originating in a node outside their cohort. On the other hand, in our approach, we rewire intra-bubble edges randomly which leads to a well-connected network for $K = 1$, where every node is highly susceptible (See Figure \ref{fig:MICU_contact_network}). However, as the value of $K$ increases, our bubbles lead to $K$ nearly disjoint components in the resulting graph. This reduces possible infection pathways, which ultimately leads to fewer infections. Finally, we also observe that we require a higher number of bubbles to reduce the infection below the baseline in MICU than in LTCF. This observation too can be attributed to natural cohorting in the MICU visit graph. Nurses in MICU show a stronger cohorting behavior which in turn leads to lower mean infection in the baseline simulation, while the nurses in LTCF cohort to a lesser degree. 

In Figure \ref{fig:infection_box_plot_unbounded_cost}, we present the distribution of infection count across 500 replicates for different values of infectivity $\rho$. We note that the bubble clustering approaches consistently result in less mean infection counts than the baseline simulation. Furthermore, \ourmethod outperforms the \random consistently. The difference in the performance gets more pronounced as transmission probability increases.

\begin{figure}[t!]
    \centerline{\includegraphics[width=0.5\textwidth]{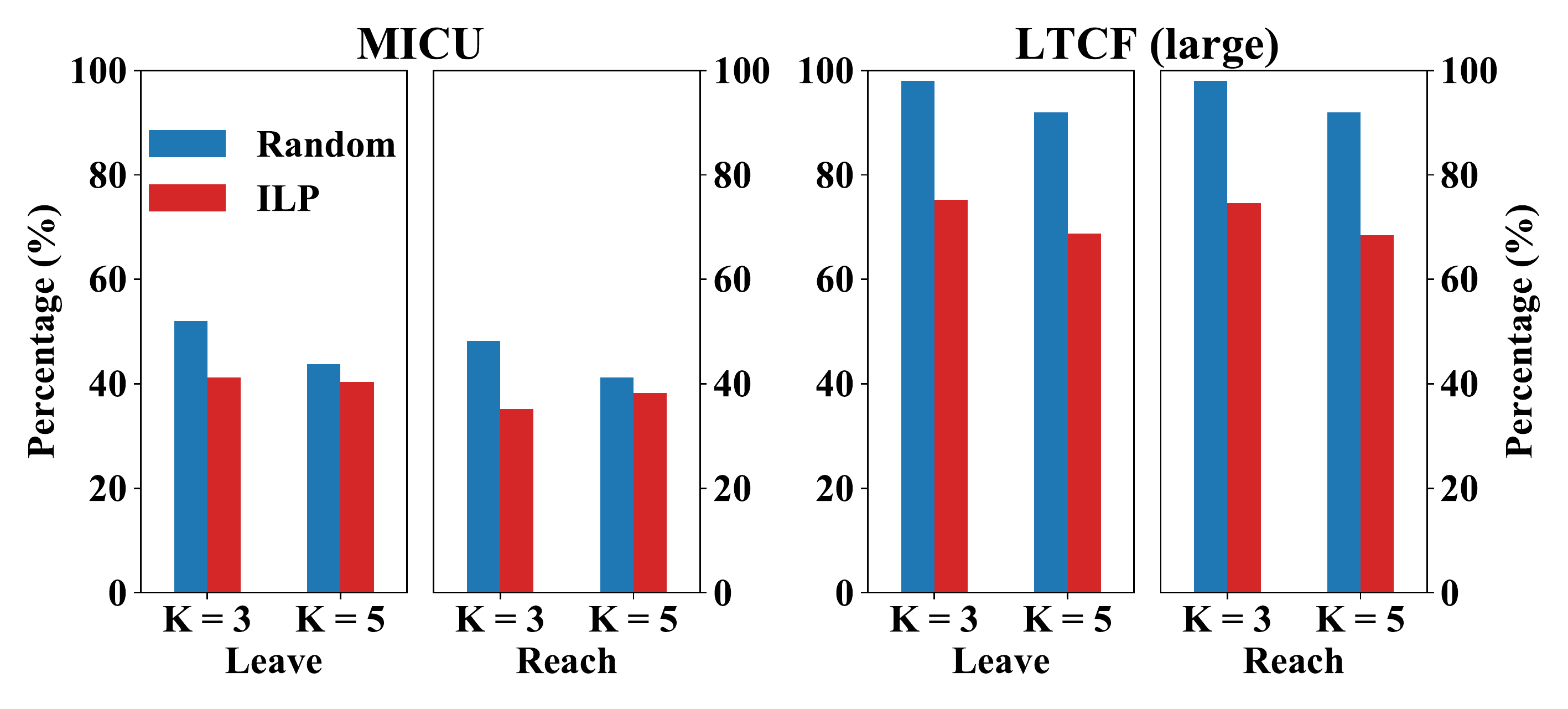}}
    \caption{
    Percentage of external transmission for different number of bubbles in the MICU and an LTCF (large) for infectivity $\rho=10 \times 10^{-4}$ ($R_0 = 2.86$) over 500 replicates.
    The $leave$ plot shows the percentage of replicates the infection leaves the source bubble and the $reach$ plot shows the fraction of $leave$ cases generating secondary infection in at least one non-source bubble.
    }
    \label{fig:external transmission}
\end{figure}

To better understand the difference in the performance of \ourmethod and \random, we also tracked the number of replicates in which the infection \textit{left} the originating bubble and those in which it \textit{reached} at least one other bubble. These results are presented in Figure \ref{fig:external transmission}. Our results indicate that the infection leaves from the originating bubble and reaches at least one other bubble in up to 26.97\% fewer replicates for \ourmethod  than that for \random. This observation explains why our approach consistently outperforms \random. 

\subsection{The cost of bubble clustering}

\begin{table}[b!]
\caption{
Unmet demand (in hours per day per room) for bubbles generated by \ourmethod, without placing any bounds on cost.
}
\centerline{
\begin{tabular}{|c|c|c|c|}
\hline
 &  \multirow{3}{*}{K} & \multicolumn{2}{c|}{Unmet demand}\\
\cline{3-4}
 &  & Average & Median \\
\hline
\multirow{2}{*}{MICU} 
 & 3 & 0.0015, 0.11\% & 0 \\
 & 5 & 0.0145, 1.11\% & 0 \\

\hline
\multirow{2}{*}{\shortstack[l]{LTCF\\(small)}} 
 & 3 & 0.0535, 4.50\%  & 0.011, 0.88\% \\
 & 5 & 0.25, 21.01\% & 0 \\
\hline
\multirow{2}{*}{\shortstack{LTCF\\(large)}} 
 & 3 & 0.0449, 7.4\%  & 0 \\
 & 5 & 0.0676, 11.15\% & 0\\
 \hline
\end{tabular}
}
\label{table:unmet_demand_ilp_without_cost_minimization}
\end{table}

Recall that some of the edges $\{ p,l\} \in E$ in the visit graph $G$ could not be rewired because HCP $p$ may not have an appropriate substitute once the bubbles have been constructed. The set of edges that were not rewired contribute towards the unmet demand of locations. While our \bc problem asks us to keep unmet demand low, we do not explicitly bound unmet demand for reasons discussed in Section \ref{sec:ipldetails}. Here we demonstrate that  unmet demand is negligible, even though it is not explicitly bounded by our ILP. We present average and median unmet demand in bubbles generated by \ourmethod for different settings in Table \ref{table:unmet_demand_ilp_without_cost_minimization}. We observe that unmet demand is fairly low in most settings with a median of 0 in most cases. A relatively higher average value for LTCF (small) $K=5$ is because there are only 8 nurses in the data. This leads to 2 bubbles with only 1 nurse each and so unmet demand is understandably higher. 

On the other hand, we explicitly bound excess load and excess footsteps in our ILP formulation. However, the results in the previous sections were obtained with the upper bound parameters $D^*$ and $Y^*$ set to $\infty$. While the unrestricted upper bound on the diameter of the bubbles and excess load allows our framework to better minimize the infections count, it may lead to high cost in terms of excess load (extra patient-care some HCPs will have to provide) and excess footsteps (extra distance HCPs will have to travel). 
High cost in terms of these two metrics could render our solution inapplicable. Here, we tabulate excess load and footsteps induced by the bubbles generated by \ourmethod for all three datasets and for $K \in \{3,5\}$ with both $D^*$ and $Y^*$ set to $\infty$ (See Table \ref{table:excess_load_and_mobility_ilp_without_cost_minimization}). 

We measure excess load in terms of hours per day per HCP and excess footsteps in meters per day per HCP.
As we see in the table, the median excess load gets as high as 3.45 hours per day per HCP. Similarly, the median excess footsteps rise to nearly 4466.1 meters. Clearly, the results show that these costs make the bubbles undesirable in actual healthcare facilities. Now a natural question that arises is whether we can effectively bound excess load and excess footsteps while maintaining low infection counts. We answer this  next.

\begin{table}[b!]
\caption{
The average and median excess load and footsteps for ILP without bounded cost.
The excess load is measured in hours per day per HCP and the excess footsteps in meters per day per HCP.
}
\centerline{
\begin{tabular}{|c|c|c|c|c|c|}
\hline
 & \multirow{2}{*}{K} &\multicolumn{2}{c|}{Excess load} & 
    \multicolumn{2}{c|}{Excess footsteps}\\
\cline{3-6}
 &  & Average & Median &  Average & Median\\
\hline
\multirow{2}{*}{MICU} 
& 3 & 0.262 & 0.523 & 234.75   & 335.97\\
& 5 & 0.211 & 0.437 & 197.94   & 279.93\\

\hline
\multirow{2}{*}{\shortstack[l]{LTCF\\(small)}} 
 & 3 & 0.529  & 1.37 & 1355.64  & 4466.1\\
 & 5 & 0.246  & 3.45 & 86.19   & 209.37 \\
\hline
\multirow{2}{*}{\shortstack{LTCF\\(large)}} 
 & 3 & 0.470 & 1.40 & 799.62 & 518.94\\
 & 5 & 0.500 & 0.577 & 518.49 & 408.51\\
 \hline
\end{tabular}
}
\label{table:excess_load_and_mobility_ilp_without_cost_minimization}
\end{table}

\begin{table}[b!]
\centering 
\caption{
The excess load (hours per day per HCP) and footsteps (meters per day per HCP) for bounded cost.
$D^*$ and $Y^*$ are also given in meters and hours respectively.
}
\begin{tabular}{|c|c|c|c|c|c|c|c|}
\hline
& \multicolumn{3}{|c|}{Parameter} &\multicolumn{2}{|c|}{Excess load} & 
    \multicolumn{2}{c|}{Excess footsteps} \\
\cline{2-8}
& $K$ & $D^*$ & $Y^*$ & Average & Median & Average & Median \\
\hline
 \multirow{2}{*}{MICU} 
 & 3 &  15 & 0.17 & 0.174 & 0.665 & 122.7 & 221.9 \\
 & 5 &  15 & 0.17 & 0.151 & 0.350 & 191.0  & 360.2\\
 \hline
 \multirow{2}{*}{\shortstack[l]{LTCF\\(small)}} 
 & 3 & 30 & 0.17 & 0.159 & 0.609 & 1215  & 2518\\
 & 5 &  15 & 0.17 & 0.171 & 0.778 &  271.6  & 776.8\\
\hline
 \multirow{2}{*}{\shortstack{LTCF\\(large)}}
 & 3 &  15 & 0.17 & 0.160 & 0.609 & 681.7 & 431.5\\
 & 5 &  15 & 0.17 & 0.204 & 0.327 &  213.9 & 284.1\\
\hline
\end{tabular}

\label{table:excess_load_and_mobility_bounded_cost}
\end{table}

\begin{figure}[t!]
    \begin{tabular}{c}
    \centerline{\includegraphics[width=0.45\textwidth]{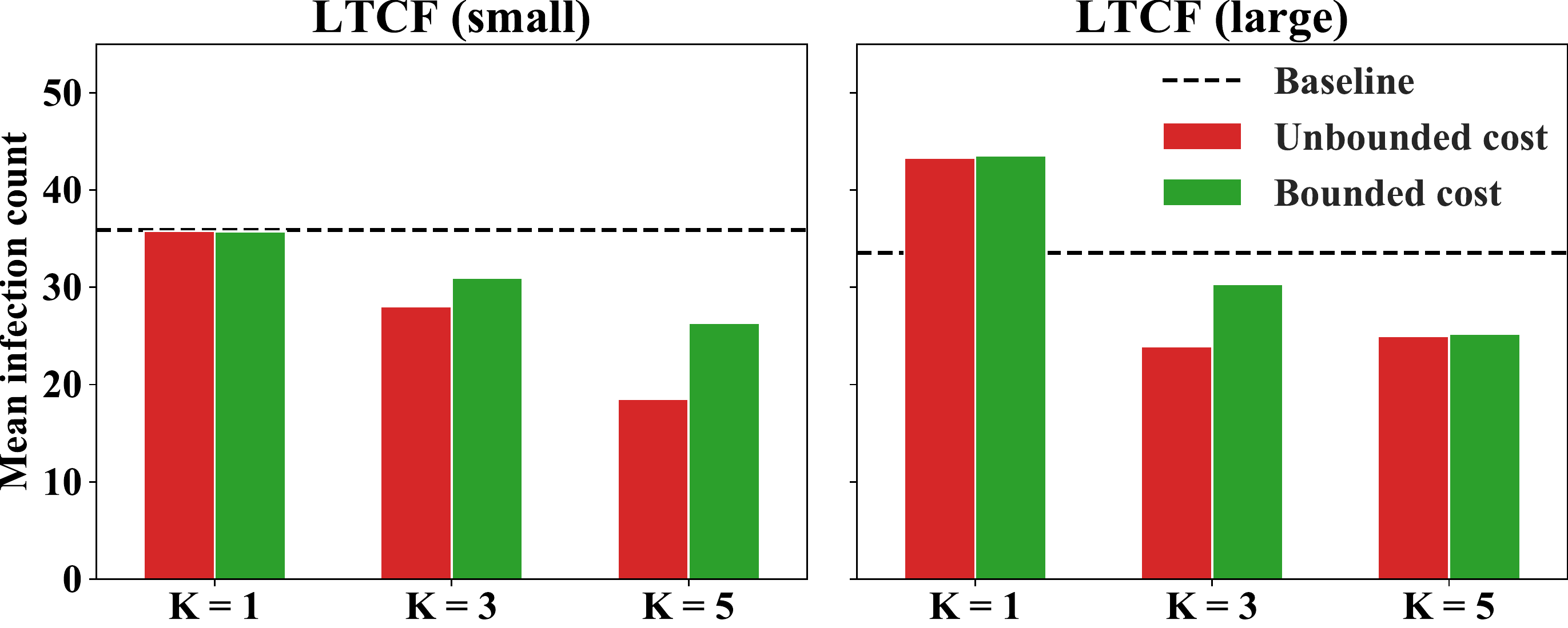}} \\
    \end{tabular}
    \caption{
    Comparison of infection counts obtained by using an ILP with unbounded costs ($D^*, Y^* = \infty$) versus bounded costs for
    $\rho = 10 \times 10^{-4}$.
    See Table \ref{table:excess_load_and_mobility_bounded_cost} for values of $D^*$ and $Y^*$ for bounded ILP.
    }
    \label{fig:comparison_bounded_ilp_and_unbounded_ilp}
\end{figure}

\subsection{Low-cost bubble clustering}

In the following, we repeat the experiments in Section \ref{subsec:infmin} for small values of the upper bound parameters $D^*$ and $Y^*$. 
Specifically, we run \ourmethod for each dataset for values of $K \in \{3,5\}$ while ranging the values of $D^*$ from $15$ to $30$ meters, keeping the value of $Y^*$ at $10$ mins ($\sim 0.17$ hrs). For each setup, we compute the average and median of both excess load and excess footsteps. Our results, summarized in Table \ref{table:excess_load_and_mobility_bounded_cost}, show that excess load and footsteps are significantly reduced.
For example, the average excess load for LTCF (small), $K = 3$ went from 3.45 hrs per day per HCP to 0.171 hrs per day per HCP.
The excess footsteps also consistently reduce.
Our results demonstrate that it is indeed possible to get a feasible bubble assignment that limits both excess load and footsteps.

Next, we study the effect of bounding excess loads and footsteps on the mean infection count. To this end, we ran a simulation on bubbles generated for each setup described in Table \ref{table:excess_load_and_mobility_bounded_cost} and computed the mean infection count. We then compare them to their unbounded counterpart. The result is presented in Figure \ref{fig:comparison_bounded_ilp_and_unbounded_ilp}. As seen in the figure, for all settings, the bubbles generated while bounding the costs maintain the infection. The average increase in infection we observed was a mere 10.02\%.

\section{Conclusions and Future Work}
\label{sec:conclusion}
In this paper, we present \ourmethod, a novel and flexible approach for solving the problem of clustering patient-care in healthcare facilities so as to minimize infection spread. 
Our approach aims to control a variety of costs to patients/residents and to HCPs so as to avoid hidden, downstream costs of clustering patient-care. We model the problem as a discrete optimization problem that we call the \bc\ problem.
Even though this problem is intractable in general, we present an ILP formulation of the problem that can be solved optimally for problem instances that arise from typical hospital units and long-term-care facilities.
Our experimental results are based on fine-grained HCP mobility data obtained from a hospital MICU and from two LTCFs. These data were obtained using sensor systems we built and deployed.
The main takeaway from our results is that it is possible to  reduce infection spread in healthcare facilities substantially by instituting clustered patient-care, while
incurring only limited costs to patients (e.g., unmet demand) and HCPs (e.g., excess load, excess footsteps).

This work is based on retrospective data. A natural next step would be to extend \ourmethod\ to solve a prospective version of the \bc\ problem in which we need to make decisions about patient-assignment to rooms, HCP assignment to patients, etc., as patients arrive/leave and patient-care demands change.

\section*{Acknowledgement}
Support for this research is provided by the Centers for Disease Control and Prevention MInD Healthcare Network (Awards U01CK000531 and U01CK000594 with respective Covid-19 supplements). 
The authors also thank other members of the Computational Epidemiology group (\verb+compepi.cs.uiowa.edu+) at the University of Iowa for helpful feedback.

\bibliographystyle{IEEEtran}
\bibliography{BIB/compepi,BIB/references}

\end{document}